\documentclass[12pt]{article}

\usepackage{amssymb}
\usepackage{amsfonts}
\usepackage{amsbsy}
\usepackage{amsmath}
\usepackage{amsthm}
\usepackage{graphicx}
\usepackage{epstopdf}
\usepackage[vcentermath]{youngtab}
\usepackage{multirow}
\usepackage{latexsym}
\usepackage{array}
\usepackage{epsfig}

\def\Z{\mathbb{Z}}
\def\F{\mathbb{F}}

\def\P{\mathbb{P}}

\def\n3a{t}

\def\tr{{\mathrm{tr}}}

\newcommand{\be}{\begin{equation}}
\newcommand{\ee}{\end{equation}}
\newcommand{\eq}[1]{eq.\ (\ref{#1})}

\newcommand {\beq}{\begin{equation}}
\newcommand {\eeq}{\end{equation}}
\newcommand {\beqa}{\begin{eqnarray}}
\newcommand {\eeqa}{\end{eqnarray}}

\begin{document}

\begin{titlepage}

\begin{flushright}
APCTP-Pre2011-002\\
OCU-PHYS345
\end{flushright}

{\begin{center} {\Large\bf  
Intersecting brane models and F-theory in six dimensions}
\end{center}
}

\begin{center}
{\large
Satoshi Nagaoka}\\
Asia Pacific Center for Theoretical Physics\\
Pohang, Gyeongbuk 790-784,
Korea\\
{\tt nagaoka} {\rm at} {\tt apctp.org}
\end{center}

\vspace*{1cm}

\abstract{We analyze six-dimensional supergravity theories coming from
intersecting brane models on the toroidal orbifold $T^4/\Z_2$.  We use
recently developed tools for mapping general 6D supergravity theories
to F-theory to identify F-theory constructions dual to the
intersecting brane models.  The F-theory picture illuminates several
aspects of these models.  In particular, we have some new insight into
the matter spectrum on intersecting branes, and analyze gauge group
enhancement as branes approach orbifold points.  These novel features
of intersecting brane models are also relevant in four dimensions, and
are confirmed in 6D using more standard Chan-Paton methods.

}

\end{titlepage}

\tableofcontents


\section{Introduction}

Six-dimensional supergravity theories provide a tractable domain in
which to explore the global structure of the space of string vacua and
the connection between different string vacuum constructions
\cite{universality, bound, KMT, tensors, 0}; a review of 6D
supergravity and string constructions can be found in
\cite{TASI-2010}.  Intersecting brane models and their T-dual magnetized
brane models \cite{Bachas,BGKL,AADS,BKL}
have been a fruitful
source of semi-realistic models of four-dimensional physics in which
many aspects of the theory can be calculated fairly easily.
Progress of the construction of vacua with Standard Model gauge group
and matter content from intersecting branes is seen in 
\cite{CSU1,CSU2,Uranga,Honecker1,HO,Honecker2,IBM-review,GBHLW,BKLS,GH}.
In this paper we consider intersecting brane
models giving rise to theories in six-dimensional space-time.  We give
a systematic description of these models on a particular toroidal
orientifold, extending earlier work on 6D intersecting brane models
\cite{bgk-6D, bbkl}.  

We revisit intersecting brane models and have here several insights 
into intersecting brane models in
the 6D context. First, we include in our analysis branes which
approach the orbifold fixed points in the compactification manifold.
This gives rise to gauge group enhancement, and a simplification of
the spectrum.  Branes away from orbifold points can be understood in
terms of a Higgsing of the theory living on the branes at the orbifold
points.  We also have some insights into the spectrum of matter
fields living at brane intersections.  In particular, we show that
additional matter fields in the adjoint representation of the gauge
group on a brane can arise when the brane intersects an orientifold
image.

These novel aspects of intersecting brane models are particularly
clear in a dual F-theory picture.  In recent work \cite{tensors}, an
explicit map was identified which takes an arbitrary 6D supergravity
theory to topological data for a corresponding F-theory construction.
Applying this map to the intersecting brane models considered here
gives an F-theory description in terms of a compactification on an
elliptically fibered Calabi-Yau threefold over a base $\F_0 =\P^1
\times \P^1$.  This correspondence turns out to be quite simple, and
relates the tadpoles associated with each factor in the gauge group of
the intersecting brane model to the topological class of the
associated divisor in F-theory.  The F-theory picture, combined with
anomaly cancellation, gives a geometric understanding 
of the matter spectrum, and sheds light on the gauge group enhancement
when branes intersect the orbifold point.

In Section \ref{sec:IBM}, we describe intersecting brane models on a
4D compactification manifold, concentrating attention on the case of
$T^4/\Z_2$.  In Section \ref{sec:F-theory} we describe the map to
F-theory, which associates topological F-theory data with the
low-energy supergravity model associated with each intersecting brane
model.  We show that part of the spectrum of the 6D theory can be
understood from geometrical structure in the F-theory picture.  In
Section \ref{sec:Chan-Paton}, we give an alternative derivation of the
spectrum using more familiar Chan-Paton methods.  In Section
\ref{sec:examples}, we give some explicit examples of intersecting
brane model theories and their F-theory counterparts.  Some concluding
discussion appears in Section \ref{sec:conclusions}.

\section{Intersecting brane models in six dimensions}
\label{sec:IBM}

Intersecting brane models are constructed by placing D-branes along
various cycles in a string compactification manifold.  The D-branes
carry world-volume gauge fields which produce the gauge group in the
space-time theory in the non-compact dimensions.  Chiral matter fields
arise from strings at the intersections between the branes.  By
choosing a compactification space which is locally flat, such as a
toroidal orbifold, the D-brane geometry becomes very simple and the
structure of the low-energy theory is relatively easy to calculate.
To preserve supersymmetry in the dimensionally reduced theory, a
negative tension object, usually an orientifold, must be included to
balance the effects of the D-branes.  In recent years, intersecting
brane models, or IBM's for short, have provided a rich supply of
easily calculable quasi-realistic models of four-dimensional string
phenomenology.  For a general review of the subject, see
\cite{IBM-review}.

Intersecting brane models in six space-time dimensions can be defined
by compactifying type IIB string theory on a K3 surface.  D7-branes
wrapping two-cycles on the K3 can be added in combination with
orientifold 7-planes to give supersymmetric string vacua.  In orbifold
limits of the form $T^4/\Z_k$, the K3 becomes locally flat and the
analysis is simplified.  We now briefly summarize the description of
intersecting brane models of this form.  Some of the structure of
these models was previously described in \cite{bgk-6D, bbkl}.  There
are, however, some differences in the brane and
matter structures we present here.  We use anomaly cancellation and
the correspondence to F-theory as a check on these formulae.
We adhere roughly to the notation of \cite{bbkl}.

\subsection{K3 as a toroidal orbifold with O7-planes}

The only topological class of Ricci-flat four manifolds other than the
four-torus is the K3 surface.  The K3 surface therefore plays a
fundamental role in supersymmetric string compactifications to six
and fewer dimensions.  There are simple toroidal orbifold limits of
K3; we focus here on the orbifold $T^4/\Z_2$.  We define $T^4 = T^2
\times T^2$ through the complex coordinates $Z_1=X_6+iX_7,
Z_2=X_8+iX_9$ subject to the
identifications
\begin{equation}
Z_i \sim Z_i + 1 \sim Z_i + \tau_i
\end{equation}
where $\tau_1, \tau_2$ are the complex modular parameters for the two
$T^2$ factors.  The $\Z_2$
orbifold giving the K3 surface is
\begin{equation}
\rho: Z_i \to -Z_i \,.
\end{equation}
This orbifold action has 16 fixed points on the $T^4$ where the
surface is singular.  Resolving these singularities by blowing them up into $\P^1$'s
gives a smooth K3.
The K3 has 22 nontrivial two-cycles, which have the intersection
lattice $\Gamma^{3, 19}$.  Six of these two-cycles descend from the
two-cycles in the $T^4$; we denote these cycles in
$H_2 (K3;\Z)$ by $\pi_{ij}$ where $i
\neq j \in\{6, 7, 8, 9\}$.
The inner product on these cycles is
\begin{equation}
\pi_{ij} \circ \pi_{kl} = -  2 \;\epsilon_{ijkl} \,,
\end{equation}
where the $\epsilon$ symbol is antisymmetric in the indices $i, j,
\ldots \in\{6, 7, 8, 9\}$.
The factor of 2 in the inner product can be seen by lifting to the
four-torus $T^4$, where a generic flat cycle $C$ has an orbifold image
$\tilde{C}$ which is disjoint from $C$ as long as $C$ does not
intersect the orbifold fixed points.  The preimage of $\pi_{ij}$ under
the orbifold action is two times the corresponding cycle
$\bar{\pi}_{ij}$ in $T^4$.  The preimages of, for example, $\pi_{67},
\pi_{89}$ intersect at 4 points, descending to 2 points in the orbifold.
Another 16 cycles come from the exceptional divisors
formed by blowing up the 16 singular points.  We denote these cycles
$e_{ij}$, where the values of the indices
$i, j \in \{1, \ldots,  4\}$
correspond to the points $(0, 0), (1/2, 0), (0, \tau/2), (1/2,
\tau/2)$ on the two toroidal factors $T^2$.
The inner product on these cycles
is $e_{ik} \cdot e_{jl} = -2 \;\delta_{ij} \delta_{kl}$.
The 22 cycles $\pi_{ij}, e_{ij}$ span a
22-dimensional sublattice (the Kummer lattice)
of the full homology lattice; to complete
the full lattice additional fractional cycles must be added, as
reviewed in \cite{Aspinwall-K3, bbkl, KT-K3}.

To include branes in a supersymmetric fashion we include orientifold
7-planes.  These are defined by imposing the discrete $\Z_2$ symmetry
$\Omega \sigma$ on the string theory, where $\Omega$ is the operator
reversing orientation on the string world-sheet, and $\sigma$ is an
isometry of space-time.  We choose
\begin{equation}
\sigma : Z_i \to \bar{Z}_i \,.
\end{equation}
This produces an orientifold 7-plane stretched along the $X_6$ and
$X_8$ directions in the compact space, as well as the 5 spatial
dimensions $X_{1-5}$.  The orbifold condition implies that the
discrete $\Z_2$ given by $\Omega \sigma \rho$ must also be a
symmetry.  This gives an additional  orientifold in the $79$
directions.  
The orientifold symmetry restricts the choice of possible complex
structures $\tau_i$ to several discrete choices; for simplicity  here
we
will take $\tau_i$ to be pure imaginary (rectangular tori).
The total homology class of the orientifold can be expressed as
\begin{align}
\pi_{O7}=2(\pi_{68}-\pi_{79}) \,,
\end{align}
The extra factor of two arises because on each transverse circle the
orientifold splits into two components, each with 1/2 the R-R charge
(this splitting
gives an overall factor of 4, which is then
divided by 2 because of the
orbifold).  For example, there is an orientifold plane along $X_7 =
X_9 = 0$, and also an orientifold plane along $X_7/R_7 = X_9/R_9 =
1/2$, where $R_{7, 9}$ are the radii of the toroidal dimensions 7 and
9.

\subsection{Branes, tadpoles, and supersymmetry}

Now let us consider wrapping D7-branes on the toroidal orbifold.  On a
simple torus $T^4$, a class of flat D7-branes can be described by
a product of 1-cycles on $T^2 \times T^2$, parameterized by
a set of
winding numbers $(n^1, m^1; n^2, m^2)$ on the compact dimensions
$X_6$-$X_9$.  Such a brane has homology class
\begin{equation}
\pi=q \; \pi_{68} -r \; \pi_{79} + s \; \pi_{69} + t \; \pi_{78} \ ,
\end{equation}
where
\begin{align}
q=n^1n^2 \ , \quad 
r=-m^1m^2 \ , \quad
s=n^1m^2 \ , \quad
t=m^1n^2 \ .
\end{align}
These coefficients satisfy
\begin{align}
qr = -st \ .
\end{align}
Note that the winding numbers on each torus must be relatively prime
for the brane to be irreducible, $(n^1, m^1) = (n^2, m^2) = 1$.

For each D7-brane with winding numbers  $(n^1, m^1; n^2, m^2)$, there
is an orientifold image brane with winding numbers
$(n^1, -m^1; n^2, -m^2)$ and charges $\pi'$ with
\begin{equation}
q' = q, r' = r, s' = -s, t' = -t \,.
\end{equation}

Both the D7-branes and orientifold planes carry charge under the R-R
8-form field dual to the IIB axion.  On a compact space, this charge must
cancel
\begin{equation}
\sum_{a} N_a (\pi_a + \pi'_a) = 8 \pi_{\rm O7} \,,
\label{eq:tadpole-o}
\end{equation}
where $N_a$ is the number of branes in a stack of identical branes
with charges $\pi_a$.
Thus, the R-R charge cancellation condition (``tadpole condition'')
given by \eq{eq:tadpole-o}
reads
\begin{eqnarray}
\sum_{a} N_aq_a  & = & 8 \label{eq:tadpole}\\
\sum_{a} N_ar_a  & = & 8   \nonumber\,.
\end{eqnarray}
We denote the contribution to this condition from a given brane type
by the pair of charges $[q, r]$ corresponding to the given brane.

For there to exist a choice of complex structures so that all branes
respect the same supersymmetry condition, we must have
\begin{equation}
n^1_a m^2_a = -K n^2_a m^1_a
\label{eq:SUSY}
\end{equation}
for some choice of (positive) $K$, for all branes $a$.  When $n^1,
n^2$ are nonzero, this is equivalent to the condition $m^2/n^2 + K
m^1/n^1 = 0$, which corresponds geometrically to the standard SUSY
condition that the branes intersect at equal and opposite angles on
the two 2-tori \cite{bdl}.  Note that \eq{eq:SUSY} implies that $q, r$
must have the same sign, so all contributions to the tadpole condition
(\ref{eq:tadpole}) are positive.

The story so far has been a summary of the basic intersecting brane
model story on a toroidal orientifold of K3, as developed in
\cite{bbkl}.

\subsection{6D gauge group and matter
spectrum from different brane types}

We now note that there are a number of distinct ways in which branes
can be wrapped.  Depending upon whether the brane is parallel to an
orientifold plane and/or intersects an orbifold fixed point, there are
four distinct possibilities for types of branes.
\vspace*{0.05in}

\noindent
{\bf (i)} Branes wrapped on orientifold cycles, intersecting orbifold
points
\vspace*{0.05in}

\noindent
{\bf (ii)} Branes parallel to orientifold cycles, not intersecting
orbifold points
\vspace*{0.05in}

\noindent
{\bf (iii)} Branes wrapped on diagonal cycles, intersecting orbifold
points
\vspace*{0.05in}

\noindent
{\bf  (iv)} Branes wrapped on diagonal cycles, not intersecting
orbifold points.
\vspace*{0.05in}

\begin{figure}
\begin{center}
\begin{picture}(200,200)(- 100,- 100)
\put(-100,40){\makebox(0,0){\includegraphics[width=8cm]{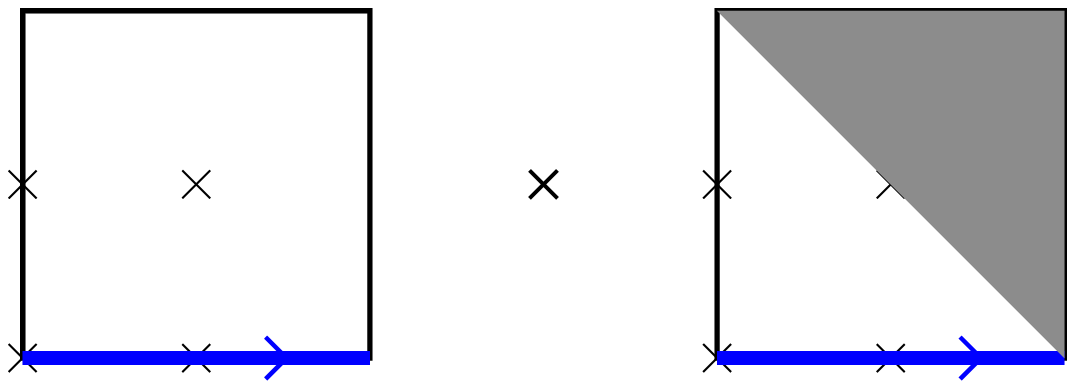}}}
\put(100,40){\makebox(0,0){\includegraphics[width=8cm]{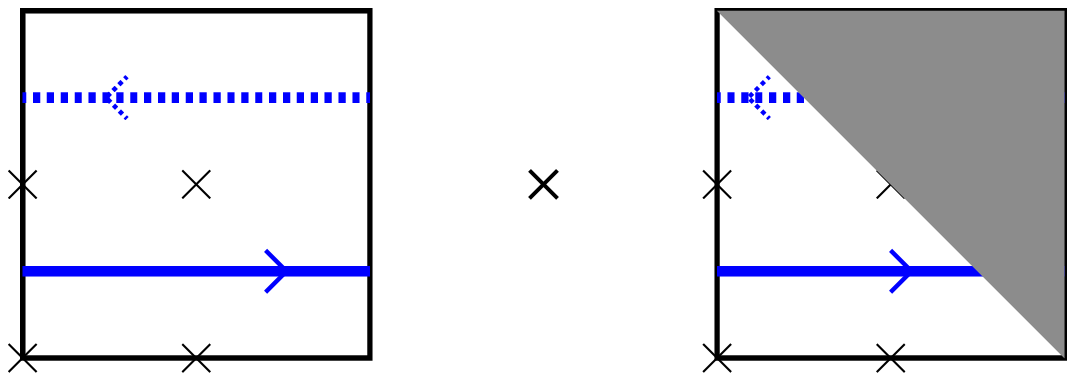}}}
\put(-100,-50){\makebox(0,0){\includegraphics[width=8cm]{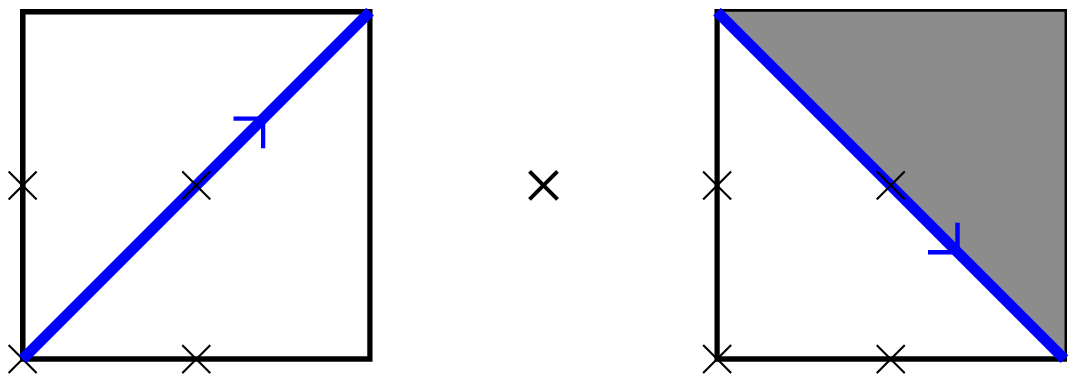}}}
\put(100,-50){\makebox(0,0){\includegraphics[width=8cm]{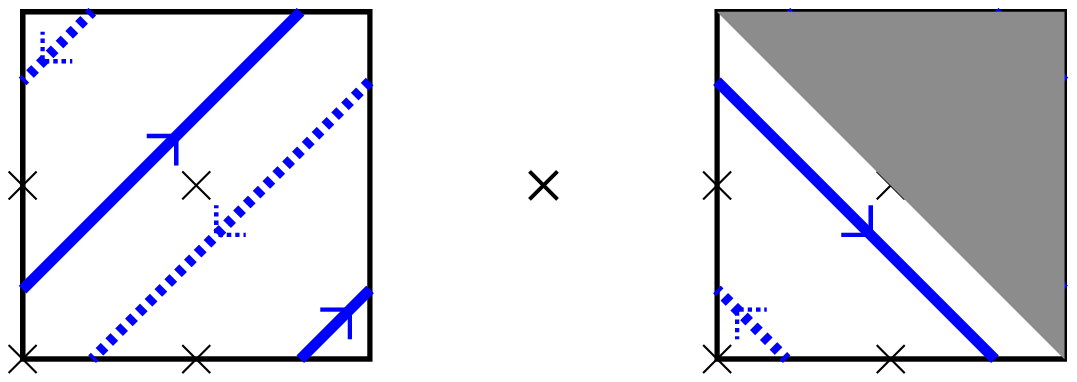}}}
\put(-100,0){\makebox(0,0){(i)}}
\put(100,0){\makebox(0,0){(ii)}}
\put(-100,-90){\makebox(0,0){(iii)}}
\put(100,-90){\makebox(0,0){(iv)}}
\end{picture}
\end{center}
\caption[x]{\footnotesize  Depiction of the four types of
  supersymmetric branes possible in  an intersecting brane model on
  the toroidal orbifold/orientifold $T^4/\Z_2$.  Branes are shown in a
fundamental domain  including only the lower left half of the second
$T^2$; dotted lines indicate the part of the brane which lifts to the
part of $T^4$ outside the fundamental domain (orbifold images). 
Note that brane types (i) and (iii) are coincident with their orbifold
  images.}
\label{f:brane-types}
\end{figure}

\noindent
A depiction of these four types of branes is given in
Figure~\ref{f:brane-types}.  

In most of the intersecting brane literature, implicitly only types
(ii) and (iv) are considered.  One of the points of this paper is to
describe how  types (i) and (iii) can be understood both in the IBM
picture and the dual F-theory picture.

We summarize the gauge group and matter spectrum arising from
these different types of branes in Table~\ref{t:table}.  
Note that all matter multiplicities in the spectrum are expressed in
terms of the ``tadpoles'' $q, r$ associated with brane charges on the
$\pi_{68}, \pi_{79}$ planes.

\begin{table}
\begin{center}
\begin{tabular}{ | r | c | c | c | c | c |}
\hline
type & $G_a$ & 
F ({\tiny\yng(1)}) &
A $\left(\raisebox{-0.1cm}[0.2cm][0.2cm]{\tiny\yng(1,1)} \right) $ &
S ({\tiny\yng(2)}) &
D \\
\hline
(i) & $SU(2N)$ &16 & 2 & 0 & 0\\
(ii) & $Sp(N)$ &16 & 1 & 0 & ---\\
(iii) & $Sp(N)$ & $16 (q + r)-4 qr N$  & $qr + q + r$ & $(q-1)
(r-1)$ & ---\\ 
(iv) & $SU(N)$ &$32 (q + r)-8q r N$ &  
$2(q +1) (r +1)$ & 
$2(q-1)(r-1)$& $2qr-1$\\
\hline
\end{tabular}
\end{center}
\caption[x]{\footnotesize Gauge group factor and matter content
  associated with different types of branes on $T^4/\Z_2$.  Matter
  includes fields in fundamental (F), antisymmetric (A), symmetric (S)
  (= $Sp(N)$ adjoint), and $SU(N)$ adjoint (D) representations.  In
  each case, the representation is actually the real representation $R +
  \bar{R}$.  We only consider nonabelian gauge group factors, ignoring
  $U(1)$ factors in the analysis of this paper.}
\label{t:table}
\end{table}

We now discuss some aspects of the origin of the spectra in
Table~\ref{t:table}.  In the next two sections we show how these
spectra can be understood in terms of the geometry of F-theory and by
standard Chan-Paton analysis.

Type (iv) branes are the most familiar type of brane appearing in
these models.  Locally, type (iv) branes lie in the bulk of the K3
space, away from the orientifold and orbifold singularities.  A stack
of $N$ such branes generally carries a gauge group $U(N)$.  Generally,
however, the $U(1)$ factors in these theories can be anomalous and
acquire a mass through the Stueckelberg mechanism.  A full treatment
of abelian factors is somewhat subtle.  We will ignore abelian factors
in the gauge group here.  We thus only concern ourselves with the
nonabelian part $SU(N)$ of the spectrum on type (iv) branes.    

As argued in \cite{bbkl}, at intersection points $x$ between $\pi$ and
the orientifold image $\pi'$, if $x$ is invariant under the
orientifold action $\sigma$ (or $\rho \sigma$) then the intersection
point carries only an antisymmetric representation.  Intersections
between $\pi$ and $\pi'$ away from an orientifold plane are more
subtle.  Depending upon the geometry of the intersection, the matter
representation may be a symmetric plus an antisymmetric
representation, or it may be a single multiplet in the adjoint
representation.  A detailed description of how these possibilities
occur in the F-theory picture is given in \cite{Morrison-Taylor}.  The
type (iv) matter spectrum given in Table~\ref{t:table} is like that
given in \cite{bbkl}, except that $2qr-2$ symmetric plus antisymmetric
representations have been replaced by additional adjoint matter
fields.  In the next section we give a simple argument from the
F-theory point of view demonstrating that the distribution of adjoint,
symmetric, and antisymmetric representations from such intersections
is that given in Table~\ref{t:table}.  The source of the additional
adjoints, which are not generally appreciated in the D-brane
literature, can be seen from the geometry of the corresponding
F-theory picture.  Type (iv) branes without their orientifold images
are topologically genus 1 curves in the K3 surface, with vanishing
self-intersection $\pi \circ \pi = 2g-2 = 0.$ Since they are genus 1
curves they carry an adjoint representation.  In the F-theory picture,
the orientifold image becomes part of the brane, and the intersection
between the brane and orientifold image augment the genus of the
brane, giving rise to additional adjoint representations.

\begin{figure}
\begin{center}
\includegraphics[width=12cm]{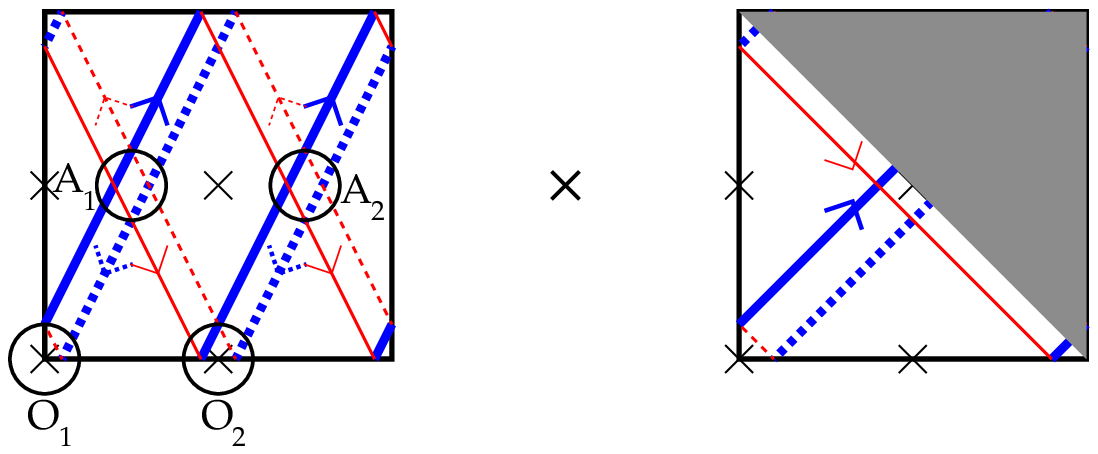}
\end{center}
\caption[x]{\footnotesize }
\label{f:enhancement}
\end{figure}

Branes of type (iii) can be seen as arising from a limit as a stack of
$N$ type (iv) branes approaches the orbifold fixed point.  In this
limit, in the toroidal cover the lift of the brane stack approaches
its orbifold image.  In the orbifold, this corresponds to parts of the
brane approaching one another, so that in the limit the two stacks of
$N$ branes merge.  An example of a brane configuration near the
enhancement point is illustrated in Figure~\ref{f:enhancement}, and
described in more detail in the following section from the F-theory
point of view.
While the type (iv) branes are wrapped on a genus 1 toroidal cycle,
the limit in which the brane approaches the orbifold counterpart
corresponds to wrapping this toroidal cycle twice around a genus 0
cycle, associated with a double cover of the sphere.  This picture is
manifest in the F-theory geometry we describe in the next section.
The type (iv) branes with $SU(N)$ gauge group can be seen as arising
from a Higgsing of the type (iii) branes with $Sp(N)$ gauge group
through VEV's for two antisymmetric matter fields.  We describe this
Higgsing picture in more detail below.

We complete the discussion of the gauge groups and matter in
Table~\ref{t:table} by briefly discussing the type (i) and (ii)
branes.  Both of these branes have gauge groups and matter content
which can be determined by a string analysis using Chan-Paton factors
and orbifold/orientifold symmetries.  The $Sp(N)$ group appearing in
type (ii) branes can be thought of as the residual gauge group after
$SU(2N)$ on a stack of type (i) branes is broken by giving an
expectation value to an antisymmetric representation, splitting the
original $2N$ branes on the orientifold plane into $N$ parallel branes
and their orientifold images.
The orientifold images of type (ii) branes are located on the orbifold
images of type (ii) branes. Thus,
strings on type (ii) are not restricted by the action $\sigma \rho$. 
The remaining projection which reduces the gauge group is $\Omega$.
Standard projection of $\Omega$ with respect to $2N$ coincident branes
gives the reduction of gauge group $SU(2N) \to Sp(N)/SO(2N)$.
In the type (ii) case, the $Sp(N)$ gauge group appears.
The matter content is a single antisymmetric representation.

\subsection{Anomalies}

The condition of anomaly cancellation places stringent conditions on
6D theories of gravity + matter with ${\cal N} = 1$ supersymmetry.  We
can use these conditions to check the spectra above.
The anomaly cancellation mechanism also provides the structure needed
to identify the corresponding F-theory model, as we will describe in
Section \ref{sec:F-theory}.

For an ${\cal N} = 1$ 6D supergravity theory with one tensor
multiplet (such as the models considered here), anomalies can be
canceled
through the Green-Schwarz mechanism \cite{gsw}
when
the 8-form anomaly polynomial $I_8$  factorizes as
\begin{align}
I_8=(\tr R^2 -\sum_a \alpha_a \tr F_a^2)(\tr R^2 -\sum_a \tilde{\alpha}_a
\tr F_a^2) \ ,
\label{eq:factorize}
\end{align}
where $F_a$ is the field strength in the $a$th nonabelian factor of
the semi-simple part of the gauge
group $G = G_1 \times \cdots \times G_k$.
The vanishing of the $R^4$ part of $I_8$ implies the well-known condition
\begin{align}
\label{R4}
H-V=244 \ ,
\end{align}
where $H, V$ are the number of scalar hypermultiplets and vector
multiplets in the theory.

Using the group theory coefficients defined through
\begin{align}
\tr_R F^2&=A_R \tr F^2 \ , \nonumber \\
\tr_R F^4&=B_R \tr F^4 +C_R (\tr F^2)^2 
\end{align}
where $\tr$ denotes the trace in the fundamental representation, and
$\tr_R$ denotes the trace in representation $R$, the
anomaly factorization
conditions (\ref{eq:factorize}) can be rewritten in a simple form.
The $F^4$ term in the anomaly polynomial vanishes when
\begin{align}
B_{Adj}^a =\sum_R x_R^a B_R^a \  \,. \label{F4}
\end{align}
Factorization of the remaining part of $I_8$ implies that there must
exist
real values of $\alpha_a, \tilde{\alpha}_a$ satisfying
\begin{align}
\alpha_a +\tilde{\alpha}_a &=\frac{1}{6} \left(
\sum_R x_R^a A_R^a -A_{Adj}^a
\right) \ , \nonumber \\
\alpha_a \tilde{\alpha}_a &=\frac{2}{3}
\left(\sum_R x_R^a C_R^a -C_{Adj}^a
\right) \ , \nonumber \\
\alpha_a \tilde{\alpha}_b +\alpha_b \tilde{\alpha}_a
&=4 \sum_{R,S} x_{RS}^{ab} A_R^a A_S^b 
 \ ,
\end{align}
where 
$x_R^a$ and $x_{RS}^{ab}$ denote the number of hypermultiplets
in the representations $R$ and $(R,S)$ of gauge group factors $G_a$
and $G_b$.

For an $SU(N)$ gauge group factor with $F, A, S, D$ matter fields in
the fundamental, antisymmetric, symmetric and adjoint representations,
the $F^4$ anomaly cancellation condition reads
\begin{equation}
F = 2 N (1-D) -A (N -8) -S (N + 8) \,.
\end{equation}
It is straightforward to check that this condition is satisfied by
each of the $SU(N)$ gauge groups in Table~\ref{t:table}.
The $F^4$ condition for $Sp(N)$ is
\begin{equation}
F =  (1-S) (2N+ 8) -A (2N -8) \,,
\label{eq:sp-condition}
\end{equation}
which is satisfied by type (i) and type (iii) branes with the spectra
in Table~\ref{t:table}

The $R^4$ condition can also be checked.  For models with only
diagonal branes of type (iii) or (iv) a short calculation using the
tadpole cancellation condition shows that
including all open string matter fields and vector multiplets gives
\begin{equation}
H_{\rm  open} -V = 224\,.
\end{equation}
This matches with the expectation that 20 hypermultiplets will be
associated with the closed string sector for models of this type
(including 16 moduli for blowing up the orbifold singularities in
addition to the axiodilaton and toroidal moduli \cite{Sen-gp}).  (A
calculation of this kind was done for models with only type (iv)
branes in \cite{bbkl}.)  In general, $H_{\rm open}$ is precisely the
number of charged fields, though there are some situations in which
these numbers differ; for example, for $SU(2)$ the antisymmetric
representation is equivalent to the one-dimensional trivial
representation and not charged.  When branes of type (i) or (ii) are
included, there are also additional uncharged scalar matter multiplets
associated with open strings.

Using the values $A_F = 1, A_{\rm Adjoint} = 2 N, A_A = N -2, A_S = N
+ 2$ and $C_F = 0, C_{\rm Adjoint} = 6, C_A = C_S = 3$ for $SU(N)$ ($N
> 3$), we can determine $\alpha, \tilde{\alpha}$ for
each gauge group factor of this kind.  The anomaly cancellation conditions 
for each such gauge group factor
become
\begin{align}
\alpha_a+\tilde{\alpha}_a&=\frac{1}{6}
\left(F_a+A_a(N_a-2)+S_a(N_a+2)-(1-D_a)2N_a
\right) \ , \nonumber \\
\alpha_a \tilde{\alpha}_a& =\frac{2}{3}\left(
3 A_a +3 S_a-6(1-D_a)
\right) \ .
\end{align}
For type (i) branes we have simply
\begin{eqnarray}
\alpha + \tilde{\alpha} & = &  2\\
\alpha \tilde{\alpha} & = &  0
\end{eqnarray}
so up to ordering we have $(\alpha, \tilde{\alpha}) = (0, 2)$.
The same result holds for type  (ii), where $S$ is replaced by $(1-D)$
in the equations for $\alpha, \tilde{\alpha}$ and the $SU(N)$ adjoint
term is absent.

For diagonal type (iv) branes, a short calculation gives
\begin{align}
\alpha+\tilde{\alpha}&=4(q + r) \ ,  \label{eq:aa} \\
\alpha\tilde{\alpha}&=16qr \ . \nonumber
\end{align}
So
\begin{equation}
(\alpha, \tilde{\alpha}) = (4q, 4r) \;\;\;\;\;
\;\;\;\;\; {\rm (type\ iv)} \,.
\label{eq:aa-4}
\end{equation}
Although the above formulae for $A_R, B_R, C_R$ are not valid for
$SU(2)$ and $SU(3)$, which have no fourth order invariant, the correct
anomaly conditions for these groups also give the result
(\ref{eq:aa-4}).
Thus, we see that the anomaly coefficients $\alpha, \tilde{\alpha}$
have simple expressions in terms of the tadpoles $q, r$ for the
diagonally wrapped branes
up to ordering.  The values in \eq{eq:aa} for a type (iv) brane become
\begin{equation}
(\alpha, \tilde{\alpha}) = (2q, 2r) \;\;\;\;\;
\;\;\;\;\; {\rm (type\ i,\ iii)} \,
\label{eq:aa-3}
\end{equation}
for type (i) and type (iii) branes.

It was shown in \cite{tensors} that in fact, with the correct
normalization for different types of gauge group, $\alpha,
\tilde{\alpha}$ are always integers and determine an integral lattice
of signature $(1, 1)$.  We will use this structure to determine
corresponding data for an F-theory model in Section
\ref{sec:F-theory}.

\subsection{Example: One stack of branes}
\label{sec:example-1}

We now give some explicit examples of complete models of intersecting
branes satisfying the supersymmetry and tadpole conditions.

The simplest solution to the tadpole constraints is to take a single
stack of $N = 8$ diagonal type (iv) branes with
\begin{equation}
(n^1, m^1; n^2, m^2) = (1, 1; 1, -1) \,.
\end{equation}
These branes have tadpole charges
\begin{equation}
[q, r] = [1, 1] \,,
\end{equation}
so this single stack satisfies the tadpole conditions
(\ref{eq:tadpole}) with no further branes added.

The gauge group and matter content for this model  are
\begin{equation}
G = SU(8), \;\;\;\;\;
{\rm matter} = 8 \times A \; ({\bf 28}) + {\rm adjoint} \; ({\bf 63}) \,.
\label{eq:model-1}
\end{equation}
It is easy to confirm that this satisfies all anomaly conditions and
has $H_{\rm charged} -V = 8 \times 28 = 224$.

Now, consider the limit where this brane stack approaches the orbifold
point.  In this limit, we have an $Sp(8)$ gauge group on 8
type (iii) branes, so that
the gauge group and matter content of the theory are
\begin{equation}
G = Sp(8), \;\;\;\;\;
{\rm matter} =  3 \times A \;({\bf 119}) \,.
\label{eq:model-1b}
\end{equation}
Again, this model satisfies the anomaly conditions and has $H_{\rm
  charged} -V = 224$.  
The limit where the brane stack hits the
orbifold point can also be undone from the type (iii) side.  The model
(\ref{eq:model-1}) arises from the model (\ref{eq:model-1b}) by
breaking the $Sp(8)$ symmetry by turning on a VEV for some of the
charged scalar fields.  In particular, under the decomposition
\begin{equation}
SU(N) \subset Sp(N) \,,
\label{eq:}
\end{equation}
the antisymmetric representation of $Sp(N)$
  branches as \cite{Slansky}
\begin{equation}
A_{Sp(N)} \rightarrow 2A_{SU(N)} \oplus D_{SU(N)} \;
( N (2N -1) -1 \rightarrow 2 \times \left(\frac{N (N -1)}{2}\right)  + (N^2 -1))
\label{eq:}
\end{equation}
The Higgsing is performed by turning on vacuum expectation values of
two fields so that $D_{SU(N)}$ breaks to $A_{SU(N)}$, giving the
matter content \eq{eq:model-1}.
Thus, these two models live on the same moduli
space of theories and are continuously connected.

\subsection{Example: Another single-stack model}
\label{sec:example-another-1}

Another solution with a single stack is to take a
stack of $N =  4$ diagonal type (iv) branes with
\begin{equation}
(n^1, m^1; n^2, m^2) = (1, 2; 2, -1) \,.
\end{equation}
These branes have tadpole charges
\begin{equation}
[q, r] = [2, 2] \,,
\end{equation}
so this single stack again satisfies the tadpole conditions
(\ref{eq:tadpole}) with no further branes added.

The gauge group and matter content for this model  are
\begin{equation}
G = SU(4), \;\;\;\;\;
{\rm matter} =  18 \times A \; ({\bf 6}) +
2 \times  S \; ({\bf  10}) +7 \times {\rm adjoint} \; ({\bf 15}) \,.
\label{eq:model-a1}
\end{equation}
As in the previous example, we can take the limit of
type (iii) branes, where
the gauge group and matter content of the theory become
\begin{equation}
G = Sp(4), \;\;\;\;\;
{\rm matter} =   8 \times A \;({\bf 27}) 
+ 1 \times S \;({\bf 36}) \,.
\label{eq:model-a1b}
\end{equation}
Again, this model can be Higgsed back to \eq{eq:model-a1} in a similar
fashion to the first example.

\subsection{Example: Two stacks and the Gimon-Polchinski model}
\label{sec:example-2}

Consider now the model with brane content
\begin{equation}
8 \times[1, 0] + 8 \times[0,1] \,.
\label{eq:}
\end{equation}
If these are type (i) branes then the gauge group and matter content
are
\begin{equation}
G = SU(16) \times SU(16), \;\;\;\;\; {\rm matter} = 2  (A, 1) + 2 (1,
A) + (F, F).
\label{eq:}
\end{equation}
This model was also described in \cite{bbkl}, and is  the
``T-dual'' of the Gimon-Polchinski model \cite{GP}.

The symmetry can be broken by moving the branes away from the
orientifold plane.  Moving them all  as a single stack parallel to
each orientifold plane gives  the model with two
stacks of type (ii) branes
\begin{equation}
G = Sp(8) \times  Sp(8), \;\;\;\;\; {\rm matter} =   (A, 1) +  (1,
A) + (F, F),
\label{eq:}
\end{equation}
where the fundamental representation of $Sp(8)$ is 16-dimensional.

Moving the branes all apart from one another, the gauge group is
broken to
\begin{equation}
G = Sp(1)^8 \times Sp(1)^8 = SU(2)^8 \times SU(2)^8 \,.
\label{eq:}
\end{equation}
This form of the gauge group was used by Sen in \cite{Sen-gp} to
construct an F-theory realization of this model.  We return to this
connection in Section \ref{sec:F-theory}.

\subsection{Example: Another two stack model}

Another example is two stack model with 
\begin{equation}
4 \times [1, 2] + 4 \times [1, 0] \, .
\end{equation}
Diagonal type (iv) $[1,2]$ branes and type (ii) $[1,0]$ branes
give
\begin{equation}
G=SU(4) \times Sp(4) \ , \;\;\;\;
{\rm matter} = 12 \times(A,1)+(1,A)+ 4 \times(F,F)+3 \times{\rm adjoint} \,.
\end{equation}
By moving the type (ii) $[1,0]$ branes onto the orientifold plane
while fixing the type (iv) $[1,2]$ branes,
we have type (i) $[1,0]$ branes and type (iv) $[1,2]$ branes.
Gauge group and matter content are
\begin{equation}
G=SU(4) \times SU(8) \ , \;\;\;\;
{\rm matter} = 12 \times(A,1)+2 (1,A)+ 4 \times(F,F)+ 3 \times{\rm adjoint} \,.
\end{equation}
Taking the limit where the (iv) branes become type (iii) gives
\begin{equation}
G=Sp(4) \times SU(8) \ , \;\;\;\;
{\rm matter} = 5 \times(A,1)+2 (1,A)+ 4 \times(F,F)\,.
\end{equation}

\section{F-theory models corresponding to IBM's}
\label{sec:F-theory}

\subsection{Mapping IBM branes to F-theory divisors}

We now show how the intersecting brane models we have described can be
mapped to corresponding models in F-theory.  F-theory is a very
general approach to constructing string vacua in even dimensions
\cite{Vafa, Morrison-Vafa-I, Morrison-Vafa-II}.
F-theory can be thought of as a geometrization of type IIB string
theory where the axiodilaton parameterizes a torus in an extra two
dimensions.  By including 7-branes with different charges, F-theory
provides a nonperturbative extension of the type IIB theory.  For
six-dimensional space-time theories, F-theory is compactified on an
elliptically fibered Calabi-Yau threefold.  For theories with one
tensor multiplet, the base $B$
of the threefold must be a Hirzebruch
surface $\F_m$.  These surfaces are essentially  $\P^1$ bundles over
$\P^1$ with a twist indexed by  $m$.  For $m = 0$, we have $\F_0
=\P^1\times \P^1$.  

In any F-theory compactification, the nonabelian  gauge group in the
space-time theory arises from a singularity locus in the elliptic
fibration localized over  an effective  irreducible
divisor class in $H_2 (B;\Z)$.
A basis for the set of
divisors in $\F_m$  is given by $D_v, D_s$, with
intersection pairings
\begin{equation}
D_v \cdot D_v = -m, \;\;\;\;\;
D_v \cdot D_s = 1, \;\;\;\;\;
D_s \cdot D_s =  0 \,.
\end{equation}
Six dimensional F-theory compactifications
are classified in terms of features of divisor structure
of the base surface of the elliptic fibration \cite{MT2}.

To understand the geometry of the F-theory models we are interested
in, we also need the canonical class
$K$ of $\F_m$, which is given by
\begin{equation}
-K = 2D_v + (2 + m) D_s \,.
\label{eq:f-k}
\end{equation}
For $\F_m$, the effective divisors which can correspond to irreducible
curves are given by
\begin{equation}
D_v,\hspace*{0.1in} 
a D_v + bD_s, \; a \geq 0, b \geq am \,.
\label{eq:f-effective}
\end{equation}

In \cite{KMT}, it was shown that given any 6D supergravity theory with one
tensor multiplet, the topological data needed to construct an F-theory
realization of the model (if one exists) can be found from the form of
the factorized anomaly polynomial (\ref{eq:factorize}).  For each
factor $G_a$, the corresponding divisor class in a Hirzebruch surface $\F_m$
is given by
\begin{equation}
(\alpha, \tilde{\alpha}) \; \rightarrow
\;\xi =\frac{\alpha}{ 2}  (D_v + \frac{m}{2} D_s)
+ \frac{\tilde{\alpha}}{ 2}  D_s \,.
\label{eq:map}
\end{equation}
This correspondence between anomaly data and F-theory data was
generalized to models with arbitrary numbers of tensors in
\cite{tensors}.
The map (\ref{eq:map}) is defined up to an interchange of the values
$(\alpha, \tilde{\alpha})$.
All smooth toric bases supporting elliptically fibered Calabi-Yau threefolds,
which are used for F-theory constructions of 
six dimensional quantum supergravity theories,
are found in \cite{MT3}.

From the simple form of equations (\ref{eq:aa-4}) and (\ref{eq:aa-3}),
we can immediately read off the result of the map (\ref{eq:map}) when
applied to mapping the branes of an intersecting brane model to
divisor classes in F-theory.  For branes of type (iv), we have
\begin{equation}
(\alpha, \tilde{\alpha})= (4q, 4r) \; \rightarrow
\;\xi =2q  (D_v + \frac{m}{2} D_s)
+ 2r  D_s = 2qD_v + (qm + 2r) D_s  \;\;\;\;\; ({\rm type\ (iv)})\,.
\label{eq:map-4}
\end{equation}
So, for example, for a model which maps to $\F_0$, the divisor
associated with a gauge group factor coming from a brane having
tadpoles $[q, r]$ is 
\begin{equation}
\xi = 2qD_v + 2rD_s
\label{eq:map-0}
\end{equation}
For type (i) and (iii) branes the values of $\alpha, \tilde{\alpha}$
are reduced by a factor of 2, and consequently so is the resulting
divisor class
\begin{equation}
(\alpha, \tilde{\alpha})= (2q, 2r) \; \rightarrow
\;\xi =q  (D_v + \frac{m}{2} D_s)
+ r  D_s = qD_v + (qm/2 + r) D_s  \;\;\;\;\; ({\rm type\ (iii)})\,.
\label{eq:map-3}
\end{equation}
This shows that for odd $q$, any associated F-theory model must have
even $m$.
In fact, all the 6D intersecting brane models we are looking at map to
$\F_0$.  This follows from the fact that there is no supersymmetric
brane allowed with a negative tadpole charge $q$ or $r$; for models
realized on $\F_m$, the charges $(q, r) = (q, -mq/2)$ would map to the
supersymmetric divisor $D_v$ and would be allowed.

We can use equations (\ref{eq:map-4}) and (\ref{eq:map-3}) to
determine the total divisor class arising from the intersecting branes
saturating the tadpole
\begin{equation}
\sum_{a}N_a (q_a, r_a) = (8, 8)
\; \rightarrow \;
\sum_{a}\nu_a \xi_a = 16D_v + (16 + 8m) D_s  = -8K\,,
\label{eq:F-theory-tadpole}
\end{equation}
where $\nu_a$ is the multiplicity of the divisor $\xi_a$ ($N$ for type
(iv), $2 N$ for types (i), (ii), (iii)).  The Kodaira
condition in F-theory which constrains the total space of the elliptic
fibration to be Calabi-Yau is
\begin{equation}
-12K = \sum_{a}\nu_a \xi_a + Y
\end{equation}
where $Y$ is an effective divisor class associated with the part of
the discriminant locus not contributing to the
nonabelian gauge group.  From \eq{eq:F-theory-tadpole}
it follows that the residual divisor is
\begin{equation}
Y = -4K \,.
\end{equation}

From the mapping (\ref{eq:map-3})
and (\ref{eq:map-4}) we can identify the nature of the transition between
type (iii) and type (iv) branes.
In F-theory this corresponds to breaking an $Sp(N)$ to $SU(N)$ by
deforming two copies of a divisor class $\xi$ into an irreducible
divisor class $2 \xi$.  We describe some explicit examples of this
below.

\subsection{Example: One-stack model}

Consider again the one-stack model described in subsection
\ref{sec:example-1}.
The 8 diagonal type (iv) branes have $[q, r] = [1, 1]$ with gauge
group $SU(8)$.  
According to \eq{eq:map-4} we can describe this model in F-theory on
$\F_0$ by embedding the $SU(8)$ on a divisor class
\begin{equation}
\xi_8 = 2D_v + 2D_s  = -K\,.
\label{eq:1-class}
\end{equation}
The genus of a curve $\xi$ in a general surface $B$ is given by the
formula
\begin{equation}
2g-2 = K \cdot \xi + \xi \cdot \xi  \,.
\end{equation}
This clearly vanishes in this case, so $\xi_8$ is a genus 1 curve, and
has an adjoint as expected.

Now, the transition to the type (iii) brane is associated with a
change of divisor class to
\begin{equation}
\xi_{iii} = \xi_{iv}/2 = D_v + D_s \,.
\label{eq:class-16}
\end{equation}
In F-theory this can occur by deforming the generic irreducible curve
on the divisor class (\ref{eq:1-class}) to a point in moduli space
where it is reducible ($2D_v + 2D_s = 2 (D_v + D_s)$).  At this point
the gauge group is enhanced to $Sp(8)$ on the divisor class
(\ref{eq:class-16}).  Computing the genus of the curve given by
\eq{eq:class-16} we have
\begin{equation}
2g-2 = K \cdot \xi_{iii} + \xi_{iii} \cdot \xi_{iii} =
-\xi_{iii} \cdot \xi_{iii} =-2 \,.
\end{equation}
This is therefore a genus 0 curve, which does not carry an adjoint
representation.  

From the fact that the F-theory divisor on which the $Sp(8)$ brane is
wrapped has genus 0, we can immediately learn something about the
spectrum of the $Sp(8)$ model.  It was shown in \cite{0} that the
genus $g$ of the curve on which a gauge group factor $G$ from a 6D
F-theory model is wrapped satisfies
\begin{equation}
g = \sum_{R}x_R g_R \,,
\label{eq:}
\end{equation}
where $g_R$ is a ``group theory'' genus associated with each
representation $R$ of the gauge group factor $G$
\begin{equation}
g_R =  \frac{1}{12}\left(2 C_R + B_R -A_R \right)\,.
\label{eq:}
\end{equation}
Since in intersecting brane models we expect at most two-index
representations, the only possible representations of $Sp(N)$ which
can occur are the fundamental, antisymmetric, and symmetric, with
corresponding genera
\begin{equation}
g_F = g_A = 0, \; g_S = 1 \,.
\label{eq:}
\end{equation}
Thus, an $Sp(N)$ group factor wrapped on a genus $g$ curve must have
precisely $g$ matter fields in the symmetric representation.  In the
case at hand, $g = 0$ so there are only antisymmetric and fundamental
representations of $Sp(8)$,
in agreement with \eq{eq:model-1b}.

\subsection{Example: Single stack with symmetric representations}

Now consider again the model \eq{eq:model-a1}.  In this case the
tadpoles are $[q, r] =[2, 2]$, so the  divisor class in $\F_0$
carrying the $SU(4)$ gauge group is $\xi_{iv} = 4 (D_v + D_s)$.
Again, this is a double cover of the divisor class
\begin{equation}
\xi_{iii} = \xi_{iv}/2 = 2 (D_v + D_s)
\label{eq:}
\end{equation}
carrying the $Sp(4)$ gauge group.  This is a genus 1 curve, so the
spectrum has one symmetric representation, in agreement with
\eq{eq:model-a1b}.  Higgsing this model reproduces the spectrum 
\eq{eq:model-a1},
in particular determining the number of adjoint representations for
the type (iv) branes.

\subsection{Higgsing $Sp(N)$ to $SU(N)$}

More generally, we can consider a general type (iii) brane with
tadpole charges $[q, r]$.  This wraps to the cycle
\begin{equation}
\xi = qD_v + rD_s
\label{eq:}
\end{equation}
in F-theory, which is a curve of genus
\begin{equation}
2g = 2 + K \cdot \xi + \xi \cdot \xi = 2 (q -1) (r -1) \,.
\label{eq:}
\end{equation}
This fixes the number of symmetric representations of $Sp(N)$ to be
$(q -1) (r -1)$ as in Table~\ref{t:table}.  Higgsing reproduces the
number of symmetrics for the resulting type (iv) branes to be
$2(q -1) (r -1)$, again as in Table~\ref{t:table}.  Thus, we see that
the geometry of F-theory coupled with the Higgsing process
uniquely determines the distribution of symmetric and adjoint
representations for the type (iv) branes.

\subsection{Example: Gimon-Polchinski model}

Next, let us consider the example model from Section
\ref{sec:example-2}.  In this case in the type (i) picture we have 8
branes with $[q_1, r_1] =[1, 0]$ and 8 branes with $[q_2, r_2]
=[0,1]$.  Mapping to F-theory on $\F_0 =\P^1 \times \P^1$ we see from
\eq{eq:map} that these branes are mapped to the divisors $2D_v$ and
$2D_s$, which are precisely the two $\P^1$ factors in $\F_0$.  This
matches perfectly with the analysis of Sen in \cite{Sen-gp}, where he
wrote explicit Weierstrass models localizing the branes on the two
$\P^1$ factors.  Moving the branes away from the orbifold point gives
a gauge group which is a product of $Sp(k)$ factors, exactly as
described in \cite{Sen-gp}.  These $Sp(k)$ factors are associated with
$k$ pairs of branes in the Sen description, wrapped on the cycle $D_v$
or $D_s$.

\section{Chan-Paton analysis}
\label{sec:Chan-Paton}

The gauge symmetry for each type (i)-(iv) is  determined by the analysis using Chan-Paton factors.
The matter content for each type is also determined by the Chan-Paton analysis, 
except for the distinction between a single  (symmetric + antisymmetric)
representation and an adjoint representation in type (iv) which are distinguished by considering 
the Higgsing process from type (iii).
We use the notation of \cite{GP}.
We introduce a Chan-Paton matrix $\lambda_{ij}$ and general open string states $|\psi,ij \rangle$
where $\psi$ is the state of the world-sheet fields.
The operator $\Omega$ reverses the orientation on the string world-sheet.
By the action of $\Omega$, the open strings are translated into 
\begin{equation} \Omega \lambda_{ij} |\psi , ij \rangle \to \lambda_{ij} (\gamma_{\Omega})_{ii'}
|\Omega \cdot \psi, j'i' \rangle (\gamma_\Omega^{-1})_{j'j}
\end{equation}
where $\gamma_\Omega$ is some matrix.
The orientifold 7-planes are defined by imposing the discrete $\Z_2$ symmetries $\Omega \sigma$ and $\Omega \sigma \rho$.
These are stretched along the $X_0,\cdots,X_5,X_6,X_8$ directions and $X_0,\cdots,X_5,X_7,X_9$ directions, respectively.

$\Omega^2$ acts on the string world-sheet as 
\begin{equation}
\Omega^2 : |\psi , ij \rangle \to
(\gamma_{\Omega }(\gamma_{\Omega }^T)^{-1})_{ii'}
|\psi,i'j' \rangle (\gamma_{\Omega }^T
\gamma_{\Omega }^{-1})_{j'j} \ .
\end{equation}
Since this operator
acts trivially
 on the world-sheet fields,
we obtain
\begin{equation}
\gamma_{\Omega }^T=\pm \gamma_{\Omega } \ .
\end{equation}

We start from strings on type (i) branes.
R-R charge tadpole cancellation condition \eq{eq:tadpole} gives the upper bound of the size of the matrix $\lambda$.
For strings on branes with [1,0] ([0,1])
tadpole charge, we describe the Chan-Paton matrix
as $\lambda^\parallel (\lambda^\perp)$. 
Since both $\sigma$ image and $\rho$ image of branes are on branes, 
$\gamma^\parallel_{\Omega \sigma}$, 
$\gamma^\parallel_{\Omega \sigma \rho}$,
$\gamma^\perp_{\Omega \sigma}$ 
and $\gamma^\perp_{\Omega \sigma \rho}$
for strings on N coincident branes satisfy 
\begin{align} 
(\gamma^\parallel_{\Omega\sigma })^T&=+ \gamma^\parallel_{\Omega\sigma }\
 , 
\quad (\gamma^\parallel_{\Omega\sigma \rho })^T
=- \gamma^\parallel_{\Omega\sigma \rho }  \ , \nonumber \\
(\gamma^\perp_{\Omega\sigma })^T&=- \gamma^\perp_{\Omega\sigma }\
 , 
\quad (\gamma^\perp_{\Omega\sigma \rho })^T
=+ \gamma^\perp_{\Omega\sigma \rho }
\end{align} 
where the sizes of these matrices are $4N (N \leq 8)$. 
The factor
$4=2 \times 2$
is coming from the orientifold image and the orbifold image of the brane stack.
$\gamma_{\Omega \sigma}^\parallel$ and 
$\gamma^\perp_{\Omega\sigma \rho }$ are determined to be symmetric
by the tadpole cancellation condition and the symmetry of the vertex
operator for massless open string vector on each brane stack.
The $\Omega \sigma$ ($\Omega\sigma \rho $)
eigenvalues of the massless vector state on branes with [1,0]([0,1]) tadpole 
charges are $-1$ since $\gamma_{\Omega \sigma}^\parallel$
($\gamma^\perp_{\Omega\sigma \rho }$) change the orientation of the
tangent derivative $\partial_t$ in the vertex operator $\partial_t X^\mu
(\mu=0,\cdots,5)$.
$\gamma^\parallel_{\Omega\sigma \rho }$ and $\gamma^\perp_{\Omega\sigma }$
are determined by considering the vertex operators for strings 
stretching between branes with
[1,0] and [0,1] tadpole charges.
Since $\gamma_{\Omega \sigma}^\parallel$ and 
$\gamma^\perp_{\Omega\sigma \rho }$ are symmetric,
symmetry of the vertex operators requires 
$\gamma^\parallel_{\Omega\sigma \rho }$ and 
$\gamma^\perp_{\Omega\sigma}$ to be antisymmetric.
In \cite{GP}, whether $\gamma$'s are symmetric or antisymmetric is
determined by the tadpole cancellation condition and
the symmetry of vertex operators,
which is related to our case by T-duality.

The simplest supersymmetric IBM which is composed of type (i) branes is to take two stacks 
of D7-branes with each stack on each orientifold plane. 
In this case, the whole size of Chan-Paton matrix $\lambda$ is 
$64 \times 64$, which is separated into two diagonal blocks 
\begin{equation} 
\lambda= \left( \begin{array}{cc} \lambda_\parallel&  \\ &\lambda_\perp \end{array}
\right)
\end{equation}
where $\lambda_\parallel (\lambda_\perp)$ are Chan-Paton matrices for 
strings on branes with [1,0] ([1,0]) tadpole charges and
the off-diagonal blocks describe the string excitation connecting the different stack of branes.

We focus on a single stack of $N$ type (i) branes with the Chan-Paton
matrix $\lambda_\parallel$, which we denote $\lambda_\parallel \to \lambda$ from now.
By
a unitary transformation, we can take $\gamma_{\Omega \sigma},\gamma_\rho$ and $\gamma_{\Omega \sigma \rho}$ 
for $\lambda$ to be 
\begin{align} 
\gamma_{\Omega\sigma }=I_{4N} \ , \quad \gamma_{\rho }=\gamma_{\Omega \sigma \rho }
= \left( \begin{array}{cc} 0& iI_{2N} \\ -iI_{2N} &0 \end{array}
\right) \ .
\end{align}
The spectrum of Neveu-Schwarz (NS) sector excited by $\psi^\mu$ with the
Chan-Paton matrix is
\begin{equation} 
\psi^\mu_{-1/2} |0,ij \rangle \lambda_{ij} \ .
\end{equation} 
The $\Omega \sigma$ 
eigenvalue of this state is
 $-1$ since $\gamma_{\Omega \sigma}$
changes the sign of the vertex operator $\partial_t X^\mu$.
The $\rho$ eigenvalue of this state is 1.
Then, the Chan-Paton matrix satisfies 
\begin{equation} 
\lambda=+\gamma_{\rho } \lambda
 \gamma_{\rho }^{-1} \ , 
\quad \lambda =- \gamma_{\Omega \sigma } \lambda^{T} \gamma_{\Omega \sigma  }^{-1} \ .
\label{eq:vec-1}
\end{equation}
The spectrum of NS sector excited by $\psi^a$ is \begin{equation}
 \psi^a_{-1/2} |0,ij \rangle 
\lambda_{ij} 
\end{equation} where $a=6,\cdots,9 $.
The $\Omega \sigma$ eigenvalue of this state is $-1$ 
since $\gamma_{\Omega\sigma }$ changes the orientation of the
tangent derivative $\partial_t$ in the vertex operator $\partial_t X^a
(a=6,8)$ and the sign of $X^a$ in $\partial_n X^a (a=7,9)$.
The Chan-Paton matrix of
 matter fields satisfies 
\begin{equation} 
\lambda=-\gamma_{\rho } \lambda
 \gamma_{\rho }^{-1} \ , 
\quad \lambda =- \gamma_{\Omega \sigma } \lambda^{T} \gamma_{\Omega \sigma  }^{-1} \ .
\label{eq:sca-1}
\end{equation}
In order to solve $\lambda$ of gauge and matter fields, 
we separate them into 4 blocks, 
\begin{equation} 
\lambda= \left( \begin{array}{cc} A& B \\ C &D \end{array}
\right) 
\end{equation}
where each block is $2N \times 2N$ matrix.
The condition (\ref{eq:vec-1}) leads
\begin{equation}
A=D=-A^T \ , \quad B=-C=B^T \ .
\end{equation}
Thus, we obtain the Chan-Paton matrix of gauge fields
\begin{align}
\lambda=
\left(
\begin{array}{cc}
A& S \\
-S &A
\end{array}
\right) \ ,
\label{eq:SU(2N))}
\end{align}
where $S$ and $A$ are symmetric and antisymmetric blocks.

The condition (\ref{eq:sca-1}) leads
\begin{align}
A=-D=-A^T \ , \quad B=C=-B^T \ . 
\end{align}
Thus, the Chan-Paton matrix of matter fields is
\begin{align}
\lambda= 
\left(
\begin{array}{cc}
A_1& A_2 \\
A_2 &-A_1
\end{array}
\right) \ ,
\label{eq:anti2}
\end{align}
where $A_1$ and $A_2$ are antisymmetric representations.
The $\lambda$ in (\ref{eq:SU(2N))}) represents $U(2N)$ gauge group and
the $\lambda$ in (\ref{eq:anti2}) is composed of 2 antisymmetric representations.
Branes with their images are classified into $(1,\Omega \sigma, \rho,
\Omega \sigma \rho)$.
In this basis, each block corresponds to the branes $(1,\Omega \sigma)$ and
$(\rho,\Omega \sigma \rho)$, respectively.
The number of fundamental representations,
which is determined by the total number of tadpoles $\sum r=8$, is 16.
Thus, the gauge group and matter content for type (i) in Table~\ref{t:table}
have been derived from the Chan-Paton analysis.
The gauge group $U(16) \times U(16)$ is obtained to take
two stacks of 16 branes on each orientifold plane,
which
is the Gimon-Polchinski model \cite{GP}.

The same argument can be applied to
the Chan-Paton factor for type (ii) [1,0] brane.
We can separate the Chan-Paton matrix for a single stack of branes 
into two diagonal blocks which act on 
branes and their $\Omega \sigma$ orientifold images respectively 
since they are located separately.
The whole Chan-Paton matrix is separated into
\begin{align}
\lambda_{\text{whole}}=\left(
\begin{array}{cc}
\lambda& 0 \\
0&\lambda
\end{array}
\right)
\end{align}
where each block whose size is $2N \times 2N$ can be regarded as the branes 
$(1,\Omega \sigma \rho)$ and $(\rho,\Omega \sigma)$, respectively.
Since the $\Omega \sigma \rho$ image of brane stack is on the brane stack, 
the orientifold action $\Omega \sigma \rho$ is written as
\begin{align}
\Omega \sigma \rho : \left(
\begin{array}{cc}
\lambda& 0 \\
0&\lambda
\end{array}
\right)_{ij} |\psi,ij \rangle \to
\left(
\begin{array}{cc}
\lambda& 0 \\
0&\lambda
\end{array}
\right)_{ij}
(\gamma_{\Omega \sigma \rho })_{ii'} | \Omega \sigma \rho \cdot
\psi,j'i' \rangle
(\gamma^{-1}_{\Omega \sigma \rho })_{j'j} \ .
\label{eq:type-ii}
\end{align}
By the standard derivation of gauge group and matter content for
unoriented strings,
we can derive those on type (ii) brane from (\ref{eq:type-ii}).
The Chan-Paton matrix for the gauge fields satisfies the condition
\begin{equation}
\lambda =-\gamma_{\Omega \sigma \rho} \lambda^T 
\gamma_{\Omega \sigma \rho}
\label{eq:type-iiCP}
\end{equation}
where $\gamma_{\Omega \sigma \rho}$ is taken as
\begin{align}
\gamma_{\Omega \sigma \rho}=
\left(
\begin{array}{cc}
0& iI_{N} \\
-iI_{N}&0
\end{array}
\right) \ .
\end{align}
Thus, we obtain $Sp(N)$ gauge group from (\ref{eq:type-iiCP}).
The Chan-Paton matrix for the matter fields satisfies
\begin{equation}
\lambda =-\lambda^T \ ,
\label{eq:anti1}
\end{equation}
which gives a single antisymmetric representation.
The solutions of Chan-Paton matrices are
\begin{align}
\lambda_{\text{whole}}=\left(
\begin{array}{cc}
A& 0 \\
0&A
\end{array}
\right) \ .
\end{align}
By counting the number of intersections with other stacks of branes,
the number of fundamental representations is obtained. It is 16 in this case.
Thus, the gauge group $Sp(N)$ and the matter content for type (ii)
seen in Table~\ref{t:table},
are derived by the Chan-Paton analysis.

The derivation of gauge group and matter content
for type (iii) branes is a bit complicated
compared to type (i) and type (ii).
After imposing the orbifold and orientifold symmetry which
are the same as in the previous argument,
we need to impose monodromy condition.
We separate the Chan-Paton matrix for gauge fields 
into two diagonal blocks 
which act on branes $(1,\rho)$
and
their orientifold images $(\Omega \sigma,\Omega \sigma \rho)$
respectively
\begin{align}
\lambda_{\text{whole}}=
\left(
\begin{array}{cc}
\lambda& 0 \\
0&\lambda
\end{array}
\right) \ ,
\end{align}
where the whole size of the matrix is $4N \times 4N$.
In this case, the orbifold image of brane stack is on the brane stack.
The orbifold action $\rho$ acts on the strings as
\begin{align}
\rho : \left(
\begin{array}{cc}
\lambda& 0 \\
0&\lambda
\end{array}
\right)_{ij} |\psi,ij \rangle \to
\left(
\begin{array}{cc}
\lambda& 0 \\
0&\lambda
\end{array}
\right)_{ij}
(\gamma_{\rho })_{ii'} | \rho \cdot
\psi,i'j' \rangle
(\gamma^{-1}_{\rho })_{j'j} \ .
\end{align}
Thus, we obtain $SU(2N)$ symmetry for the gauge group.
The whole Chan-Paton matrix is described as
\begin{align}
\lambda_{\text{whole}}=
\left(
\begin{array}{cc}
D& 0 \\
0&D
\end{array}
\right) \ 
\end{align}
where $D$ is a hermitian block.
However, since type (iii) brane should be smoothly connected to 
type (iv) brane in the moduli space of theories,
as an additional constraint, we need to impose a monodromy condition.
Then, the Chan-Paton matrix 
satisfies
\begin{equation}
\lambda =\gamma \lambda^T 
\gamma
\label{eq:monodromy}
\end{equation}
where $\gamma$ is taken to be
\begin{align}
\gamma=
\left(
\begin{array}{cc}
0& I_{N} \\
I_{N}&0
\end{array}
\right) \ .
\end{align}
After all, we obtain S$p(N)$ as a gauge group for type (iii) brane.
The matter fields in the fundamental representation are obtained
when branes hit another branes. For N type (iii) branes with
tadpole charges [q,r],
the number of fundamental representations are counted as $16(q+r)-4qrN$.
The symmetric and antisymmetric matter representations are 
obtained when branes hit their orientifold images.
Note that since type (i) and (ii) branes are on their orientifold images,
they have no intersections with their orientifold images.
As we have shown previously,
the matter contents for type (i) and (ii) are determined in the same way
as the gauge groups.

The intersections between type (iii) branes and their orientifold images are
classified into three cases.

\noindent
{\bf (I)} The intersection points located on orbifold points

\noindent
{\bf (II)} The intersection points located away from orbifold points, but on
orientifold planes

\noindent
{\bf (III)} The intersection points located away from orientifold planes

For the case (I), the intersection points are not moved by 
the actions of orbifold $\rho$ and orientifold $\sigma$. 
The symmetry of the fields at these points 
is the same as that of matter fields on type (i) branes.
Thus, matter fields coming from each point of the case (I) are 2
antisymmetric representations, which can be seen from (\ref{eq:anti2}).
Any type (iii) models have two intersection points in case (I).
By imposing the monodromy condition (\ref{eq:monodromy}) to the Chan-Paton
matrix (\ref{eq:anti2}), a single antisymmetric representation vanishes
from one of the two intersection points.
In all, three antisymmetric representations are given by the case (I) 
intersections.

For the case (II), the intersection points are moved by the action of
orbifold $\rho$, but not moved by the action of orientifold $\sigma$.
The symmetry of the fields at these points 
is the same as that of matter fields on type (ii) branes.
Thus, matter fields coming from each point of the case (II)
are a single antisymmetric representation, 
which can be seen from (\ref{eq:anti1}).

For the case (III), the intersection points are moved either by the
action of orbifold $\rho$ or by the action of orientifold $\sigma$.
Since the branes are on their orbifold images $\rho$,
the orbifold action $\rho$ for Chan-Paton matrix is
\begin{align}
\rho :
\lambda_{ij}|\psi,ij \rangle=\left(
\begin{array}{cc}
\lambda_1& \lambda_2 \\
\lambda_3&\lambda_4
\end{array}
\right)_{ij} |\psi,ij \rangle &\to
\left(
\begin{array}{cc}
{\lambda_1}&{\lambda_2} \\
{\lambda_3}&{\lambda_4}
\end{array}
\right)_{ij}
(\gamma_{\rho})_{ii'} |\rho \cdot
\psi,j'i' \rangle
(\gamma^{-1}_{\rho})_{j'j} \ ,
\end{align}
where each block can be regarded as the branes $(1,\rho)$ and $(\Omega
\sigma \rho, \Omega \sigma)$, respectively. 
$\lambda_1$ and $\lambda_4$ do not include the intersections between
the branes and their orientifold images, so they vanish
\begin{align}
\lambda_1=\lambda_4=0 \ .
\end{align}
We need to consider pairs of intersection points which are replaced
with each other by the action of orientifold. Any intersections of type
(iii) branes away
from orientifold plane have such partners of intersection points in the
fundamental domain. These two intersection points give a symmetric and
an antisymmetric representation. 
Chan-Paton matrix for these two intersection points is
\begin{align}
\lambda_{\text{whole}}=\left(
\begin{array}{cc}
0& S+A \\
S+A&0
\end{array}
\right) \ ,
\end{align}
where a single symmetric and antisymmetric representations are obtained.
Matter fields coming from each point of the case (III) are $(S+A)/2$.
By counting the intersection numbers of these three cases, 
one can reproduce the matter content 
for type (iii) brane in Table~\ref{t:table}.

The Chan-Paton matrix for gauge fields of type (iv) brane is separated into
4 diagonal blocks since both the orientifold and orbifold images are not
on branes
\begin{align}
\lambda_{\text{whole}} =
\left(
\begin{array}{cccc}
\lambda& 0&0&0 \\
0&\lambda&0&0\\
0&0&\lambda&0\\
0&0&0&\lambda
\end{array}
\right) \ .
\end{align}
 Each block whose size is $N \times N$
is regarded as 1,$\Omega \sigma$,$\rho$ and $\Omega
\sigma \rho$, respectively. 
Off-diagonal blocks are not relevant for the massless excitations.
The solution of the whole Chan-Paton matrix is
\begin{align}
\lambda_{\text{whole}} =
\left(
\begin{array}{cccc}
D& 0&0&0 \\
0&D&0&0\\
0&0&D&0\\
0&0&0&D
\end{array}
\right) \ ,
\end{align}
where the gauge group for type (iv) model is $SU(N)$.
As we have considered pairs of intersections in type (iii) brane,
we need to consider groups of four intersections in type (iv) brane
to solve Chan-Paton matrices for matter fields.
Dynamically, these four intersections come from a single intersection of
type (iii) brane through Higgsing.
Two of them are the intersections between branes and their orientifold
image (1 and $\Omega \sigma$) and its orbifold image ($\rho$ and $\Omega \sigma
\rho$).
The other two are the intersections between (1 and $\Omega \sigma \rho$)
and ($\rho$ and $\Omega \sigma$).
Groups of four intersections are classified into three cases coming from
three cases of intersections in type (iii) brane.
The first case is coming from the case (II) in type (iii) brane.
A single brane stack is intersecting with its orientifold image
on orientifold plane which appears through Higgsing process.
This intersection gives a single
antisymmetric representation. The orbifold image of this
intersection also gives a single antisymmetric representation. 
Since the other two are not on 
orientifold plane, one of the intersections gives an adjoint
representation. These are summarized in the solution of the Chan-Paton matrix 
\begin{align}
\lambda_{\text{whole}} =
\left(
\begin{array}{cccc}
0& A_1&0&D \\
A_1&0&D&0\\
0&D&0&A_2\\
D&0&A_2&0
\end{array}
\right) \ .
\end{align}
Thus, an adjoint and two antisymmetric representations are 
obtained from a group of four intersections
of the first case. This is consistent with the decomposition
\begin{equation}
A_{Sp(N)} \rightarrow 2A_{SU(N)} \oplus D_{SU(N)} \ .
\end{equation}

The second case is coming from the case (III) in type (iii).
Any of four intersection points are not on orientifold plane in this case.
Then, two of them give a single symmetric and antisymmetric
representation and one of the other two gives an adjoint representation.
These are summarized in the Chan-Paton matrix 
\begin{align}
\lambda_{\text{whole}} =
\left(
\begin{array}{cccc}
0& S+A&0&D \\
S+A&0&D&0\\
0&D&0&S+A\\
D&0&S+A&0
\end{array}
\right) \ .
\end{align}
An adjoint and a single symmetric and antisymmetric representations 
are obtained from a group of four intersections
of the second case. 
As we have shown previously, a pair of intersections in type (iii) model
give $(S+A)$. Thus, Higgsing process is consistently reproduced since the
decompositions for the antisymmetric and symmetric representations are
\begin{align}
A_{Sp(N)} \rightarrow 2A_{SU(N)} \oplus D_{SU(N)} \ , \nonumber \\ 
S_{Sp(N)} \rightarrow 2S_{SU(N)} \oplus D_{SU(N)} \ .
\end{align}
The pair of intersection points in type (iii) are splitted into eight
intersections in all, which give $2(D+A+S)$ in type (iv) model.

The remaining case is coming from the case (I) in type (iii) brane.
All the type (iii) models include two intersection points in
the case (I). 
Two antisymmetric representations appear in each intersection of 
the case (I) in type (iii).
Chan-Paton matrices for groups of four intersections in
type (iv) brane are described as
\begin{align}
\lambda_{\text{whole}} =
\left(
\begin{array}{cccc}
0& A_1&0&D \\
A_1&0&D&0\\
0&D&0&A_2\\
D&0&A_2&0
\end{array}
\right) \ , \quad
\lambda_{\text{whole}} =
\left(
\begin{array}{cccc}
0& D&0&A_3 \\
D&0&A_4&0\\
0&A_4&0&D\\
A_3&0&D&0
\end{array}
\right) \ .
\label{eq:case-I}
\end{align}
An adjoint and four antisymmetric representations are obtained 
from a group of four intersections in this case. 
Let us remind you that 
total number of antisymmetric representations in the case (I) of type
(iii) is three, which is because
an antisymmetric representation appearing in one of the two intersection points
in type (iii) model vanishes by imposing monodromy condition.
So only one of the two intersection points in case (I) is applied to the
solution (\ref{eq:case-I}).
Chan-Paton matrix for the group of four intersections in
type (iv) brane coming from the other intersection in type (iii)
is
\begin{align}
\lambda_{\text{whole}} =
\left(
\begin{array}{cccc}
0& A_5&0&A_6 \\
A_5&0&A_7&0\\
0&A_7&0&A_8\\
A_6&0&A_8&0
\end{array}
\right) \ .
\end{align}
In all, an adjoint and eight antisymmetric representations are obtained from
this case.
This is consistent with the decomposition
\begin{equation}
A_{Sp(N)} \rightarrow 2A_{SU(N)} \oplus D_{SU(N)} \rightarrow 3A_{SU(N)}\ ,
\end{equation}
where two of the antisymmetric representations are decomposed into
$3A$ and the other one is decomposed into $D+2A$.

By counting the intersection numbers, we can reproduce the matter content
for type (iv) brane in Table~\ref{t:table}.

\section{More examples and classification of models}
\label{sec:examples}

The general classification of 6D intersecting brane models is quite
straightforward compared to the analogous problem for 4D theories.
Unlike 4D models, where some tadpoles can be negative, complicating
the classification problem \cite{Blumenhagen-statistics, Douglas-Taylor,
Rosenhaus-Taylor}, for the 6D models the tadpoles $[q, r]$ are
positive for all branes, so the classification of possible solutions
to (\ref{eq:tadpole}) amounts to a straightforward partition problem.
Furthermore, since as we have seen the brane spectra depend only on
the types of branes and the tadpoles, once we have solved the
partition problem we know the spectra of all possible models.
The remaining issues are that, on the one hand, some tadpole
combinations may not be possible to reproduce from any combination of
winding numbers for irreducible branes, and, on the other hand, some
tadpole combinations may arise from several different combinations of
brane winding numbers.  We now briefly summarize the classification of
solutions, and address these issues.

\subsection{Single stack models}

For models with only a single stack of branes there are few
possibilities.  The brane tadpoles must be proportional to $[1, 1]$
and must divide 8, so the only possible models for type (iv)
diagonal branes are
\begin{equation}
\begin{array}{rll}
8 \times [1, 1]: & \;\;\;\; G = SU(8) & \;\;\;\; {\rm matter} = 8\times A+ {\rm adjoint}\\
4 \times [2, 2]: & \;\;\;\; G = SU(4) & \;\;\;\; {\rm matter} = 18\times
 A + 2\times S + 7 \times {\rm adjoint}\\
2 \times [4, 4]: & \;\;\;\; G = SU(2) & \;\;\;\; {\rm matter} =
 18\times S + 31 \times {\rm adjoint}\\
1 \times [8, 8]: & \;\;\;\; G = {\rm abelian} & 
\end{array}
\label{eq:1-stack}
\end{equation}
In each case, a transition to type (iii) branes gives a corresponding
model 
\begin{equation}
\begin{array}{cll}
16 \times [1/2, 1/2]_{{\rm iii}}: & \;\;\;\; G = SU(16) & \;\;\;\; {\rm matter} = 3 \times
A\\
8 \times [1, 1]_{{\rm iii}}: & \;\;\;\; G = SU(8) & \;\;\;\; {\rm matter} = 8\times A + 1
\times S\\
4 \times [2, 2]_{{\rm iii}}: & \;\;\;\; G = SU(4) & \;\;\;\; {\rm matter} = 24\times A + 9\times S \\
2 \times [4, 4]_{{\rm iii}}: & \;\;\;\; G = SU(2) & \;\;\;\; {\rm matter} = 49\times S 
\end{array}
\label{eq:1-stack-3}
\end{equation}
where $[q/2 ,r/2]_{\rm iii}$ are tadpoles from fractional charges.

Let us consider a few specific aspects of these models.
In each case it is straightforward to verify that the anomaly
conditions are satisfied.
In each case the type (iv) branes can be chosen in parallel groups
which are not coincident, giving a breaking of the full gauge group.
This corresponds to turning on a Higgs VEV in the adjoint
representation to separate the branes in the transverse directions.
For example, in the first model the $SU(8)$ group can be broken to
$SU(N) \times SU(8-N)$ by splitting the branes into two groups.  The
gauge groups on type (iii) branes cannot be broken in this way since
there is no adjoint; geometrically, the fractional branes cannot be
moved away from the orbifold point.

The first  models in the lists
(\ref{eq:1-stack}),
(\ref{eq:1-stack-3}) correspond to the type (iv)
and (iii) versions of the example considered in Subsection
\ref{sec:example-1}.  This model is uniquely realized by the set of winding
numbers given there.  

The second pair of models can be realized by two different
combinations of winding numbers.  Considering the type (iv) branes,
the tadpoles $[2, 2]$ can be produced by the winding number
combinations
\begin{eqnarray}
(n^1, m^1; n^2, m^2)  &= &(1, 2; 2, -1)  \label{eq:winding-2}\\
(n^1, m^1; n^2, m^2)  &= &(2, 1; 1, -2) \,. \nonumber
\end{eqnarray}
These two models are equivalent by a rotation of axes for the torus
leaving the orbifold and orientifold action unchanged.
The other pairs of models can also be realized in two equivalent ways,
by replacing $2 \rightarrow 4, 8$ in the winding numbers in
\eq{eq:winding-2}.  There are no other possibilities because of the
condition that the winding numbers are relatively prime on each
two-torus, $(n^i, m^i) = 1.$

\subsection{Models with two or more stacks}

It is straightforward to enumerate all two-stack models.  We outline
this process and mention some relevant issues.

Considering combinations of diagonal branes, and focusing on type (iv)
branes, there are a limited set of possibilities.  The only
possibilities for solving the partition problem from the tadpole
condition for a pair of mutually supersymmetric branes
are
\begin{equation}
\begin{array}{cll}
2 \times [1, 3]+
2 \times [3, 1]: & \;\; \;G = SU(2) \times SU(2) &
\; \;\; {\rm matter} = 5 \times {\rm adjoint} + 5 \times {\rm adjoint}
\end{array}
\label{eq:2-stacks}
\end{equation}
\begin{equation}
1 \times [q, 8-q]+
1 \times [8-q, q], q \in\{1, 2, 3\}: \;\;\;\; G = {\rm abelian} 
\nonumber
\label{eq:}
\end{equation}
Note that other solutions to the partition problem
such as $[8, 8] = (1, 5) + (7, 3)$ cannot be realized by any winding
numbers which satisfy the SUSY condition (\ref{eq:SUSY}).  For
example, any prime factor in the numerator or denominator of the ratio
$K$ appearing in \eq{eq:SUSY} must appear in one of the tadpoles of
every brane in a supersymmetric model; this consideration alone rules
out most other possibilities, and the remaining possibilities can be
ruled out on a case-by-case basis.
For similar reasons, each of these models is essentially unique up to
equivalence.  For example, the winding numbers of the branes in the
first model giving tadpoles $[1, 3] +[3, 1]$ can be
$(1, 1; 1, -3) + (3, 1; 1, -1)$ but not
$(1, 1; 1, -3) + (1, 1; 3, -1)$  since the second combination violates
the SUSY condition.
In some cases
there are several distinct combinations of winding numbers 
giving equivalent tadpoles which are not obviously
equivalent under a simple relabeling of toroidal axes.
For example
for the
abelian model with tadpoles $[2, 6] +[6, 2]$, the winding numbers
could be
\begin{equation}
(2, 1; 1, -6) + (6, 1; 1, -2) \;\;\;\;\; {\rm or} \;\;\;\;\;
(1, 2; 2, -3) + (3, 2; 2, -1) \,.
\end{equation}
In the following section we show that these models map to equivalent
topological data for F-theory constructions, and are therefore
topologically equivalent; these models correspond to points in the
same moduli space of 6D theories related by a continuous deformation
of modular parameters, although this deformation may go outside the
space of intersecting brane model constructions on flat toroidal
orbifolds.

In all the models listed in \eq{eq:2-stacks}, as in the one stack
models, any stack can be split through Higgsing or moved to the
orbifold point to produce a type (iii) stack.

Now consider two-stack combinations where one stack is composed of
diagonal branes, say of type (iv), and the other stack contains
``filler'' branes of type (i/ii).  Without loss of generality we can
choose the filler branes to have tadpoles $[1, 0]$.  Since the filler
branes are parallel to the orientifold plane and automatically are
supersymmetric, we can choose the diagonal stack to contain branes
with any allowed tadpole numbers.  So we can choose any brane stack $N
\times[p, 8/N]$ with $p \leq 8/N$ for the diagonal branes, which with
$8-N p$ filler branes satisfies the tadpole conditions.  Such a
configuration is possible for any $N = 1, 2, 4$ and $p \leq 8/N$.  For
$ N = 8$ we get the combination of tadpoles $8 \times[1, 0] + 8
\times[0, 1]$ from the example above in subsection
\ref{sec:example-2}.

It is straightforward to construct in an analogous fashion all
possible models with 3 or more distinct blocks associated with branes
having different tadpole combinations.  This leads, for example, to
three types of models with stacks of sizes $3, 2, 1$:
\begin{eqnarray}
& &3 \times[1, 2] + 2 \times [2, 1]+ 1 \times [1, 0] \\
& &3 \times[1, 0] + 2 \times [2, 2]+ 1 \times [1, 4]  \nonumber\\
& &3 \times[1, 0] + 2 \times [0, 1]+ 1 \times [5, 6] \nonumber \,.
\end{eqnarray}
These models have nonabelian gauge group $SU(3)\times SU(2)$ in
addition to possible abelian factors.

\section{Conclusions}
\label{sec:conclusions}

We have analyzed ${\mathcal N}=1$ six dimensional supergravity
theories which are coming from intersecting brane models.
Intersecting brane models in six dimensions are defined by
compactifying type IIB superstring theory on K3 surface with 
D7-branes wrapping two cycles on K3.
By taking toroidal orbifold limits of K3, we obtain the orbifold
$T^4 / \Z_2$. Branes can be wrapped in various ways.
We have analyzed four types of branes which are classified by
whether the brane is parallel to an orientifold plane and/or
intersects an orbifold fixed point.
Matter fields arise from strings at the intersections between
branes. Gauge groups and matter spectrum have been clarified 
for each type of branes.
In particular, we have had some insights into the matter 
spectrum on intersecting branes approaching the orbifold fixed points
(iii) and away from orbifold fixed points (iv).
The type (iv) branes can be seen as arising from the Higgsing of the 
type (iii) branes through the VEV for antisymmetric matter fields.
These gauge groups and spectrum have been also derived from Chan-Paton methods.

We have described the map from intersecting brane models to F-theory.
Tools for mapping six dimensional supergravity model to 
F-theory topological data are used to identify F-theory constructions
dual to intersecting brane models.
Our results of the mapping of type (i) and (ii) brane
agree with Sen's results \cite{Sen-gp,Sen}.
The mapping of type (iii) and (iv)
includes the generalization of Sen's analysis.

Much of the structure described here has similar analogues for
compactifications to four dimensions. In \cite{GT}, 
Green-Schwarz like axion-curvature squared terms are used 
to identify F-theory compactifications from supergravity
with analogous gauge group and matter structure.
It is interesting to analyze six dimensional supergravity 
theories coming from intersecting brane models on other orbifolds 
$T^4/\Z_N$ \cite{Nagaoka}.

\vspace*{0.1in}

\noindent
{\bf Acknowledgements}: 
We would like to thank Washington Taylor for 
the collaboration in parts of this project.
We would also like to thank Vijay Kumar and 
Jaemo Park for helpful discussions.  
This research was
supported 
in part by 
the JSPS Institutional Program for
Young Researcher Overseas Visits
" Promoting international young researchers in mathematics and
mathematical sciences led by OCAMI ". 

\appendix

\end{document}